\documentclass[10pt,a4paper]{amsart} 
\usepackage{amsfonts,amssymb,amsthm,ifpdf,dsfont,mathrsfs,esint}
\usepackage{enumitem} 
\usepackage{kbordermatrix}

\usepackage{tikz}
\usetikzlibrary{arrows,positioning,decorations.pathmorphing,
  decorations.markings} 

\tikzset{inner sep=0pt, 
  root/.style={circle,draw,minimum size=7pt,thick}, 
  cross/.style={path picture={ 
  \draw[black]
(path picture bounding box.south east) -- (path picture bounding box.north west) (path picture bounding box.south west) -- (path picture bounding box.north east);
}},
  aio-root/.style={circle,fill,minimum size=7pt}, 
  io-root/.style={draw,circle,cross,minimum size=7pt}, 
  doublearrow/.style={double,double equal sign distance,-implies}
} 

\usepackage[overload]{textcase}

\ifpdf
  \usepackage[final,expansion=alltext,protrusion=alltext]{microtype} 
\fi

\usepackage{xargs} 
\usepackage{mathtools} 
\usepackage{ifthen} 

\theoremstyle{plain}

\newtheorem*{Th*}{Theorem}

\newtheorem*{Cor*}{Corollary}

\theoremstyle{definition}

\theoremstyle{remark}

\makeatletter

\newcommand{\IfUpperCase}[1]{\begingroup 
  \protected@edef\@tempa{\expandafter\@firstofone\@firstofone#1.}%
  \expandafter\IfUpperCasE \@tempa\delimiter}

\def\IfUpperCasE #1#2\delimiter{%
  \protected@edef\@tempa{\meaning#1\meaning a}%
  \ifnum \expandafter\IfUppercaSE\@tempa \IfUppercaSE
   \endgroup \expandafter\@firstoftwo
  \else
   \endgroup \expandafter\@secondoftwo
  \fi}

\@ifundefined{strip@prefix}{\def\strip@prefix#1>{}}{}
\def\@tempa{the letter }
\edef\@tempa{\expandafter\strip@prefix\meaning\@tempa}

\expandafter\def\expandafter\IfUppercaSE\expandafter#\expandafter1\@tempa#2#3\IfUppercaSE{\uccode`#2=`#2 }

\newif\ifuc@se
\def\setuc@se#1{\IfUpperCase{#1}{\uc@setrue}{\uc@sefalse}}

\newif\ifTNS 
\TNStrue

\def\printtheoremname#1{\ifuc@se \lowercase{\csname#1name\endcsname}\ignorespaces%
  \else \edef\@temp{\lowercase{\lowercase{\csname#1name\endcsname}}}\@temp\ignorespaces%
  \fi}
\def\printtheoremnames#1{\ifuc@se \lowercase{\csname#1names\endcsname}\ignorespaces%
  \else \edef\@temp{\lowercase{\lowercase{\csname#1names\endcsname}}}\@temp\ignorespaces%
  \fi}
 
 \def\lcprinttheoremnames#1{\uc@sefalse\lowercase{\printtheoremnames{#1}}}%
 
\def\thmref#1#2{\setuc@se{#1}\lowercase{{\printtheoremname{#1}\ifTNS~\fi\lowercase{\ref{#1:#2}}}}}

\def\uc#1#2{\MakeUppercase{#1}{#2}} 

\newcommand{\DefTheorem}[2]{\newenvironmentx{#1}[2][1=\empty,2=\empty]{%
    \ignorespaces%
    \ifx##2\empty%
      \begin{#2}%
    \else%
      \begin{#2}[{\uc##2}]%
    \fi%
    \ifx##1\empty%
      {}%
    \else%
      \lowercase{\label{#1:##1}}%
    \fi%
    \ignorespaces}{\end{#2}\ignorespacesafterend}}

\makeatother

\DefTheorem{Th}{theorem}
\DefTheorem{Prop}{proposition}
\DefTheorem{Cor}{corollary}
\DefTheorem{Lem}{lemma}
\DefTheorem{Def}{definition}
\DefTheorem{Rem}{remark}
\DefTheorem{Par}{para}
\DefTheorem{Not}{notation}
\DefTheorem{Exer}{exercise}
\DefTheorem{Ex}{example}
\DefTheorem{Cons}{construction}
\DefTheorem{Scho}{scholium}

\newenvironment{Par*}{\ignorespaces\noindent\ignorespaces}{\ignorespacesafterend}

\numberwithin{equation}{section}

\newcommand\Define[2][\empty]{\ignorespaces%
  \emph{#2}}%

  {\ignorespaces\end{tikzpicture}%
  \end{minipage}\\%
  \ignorespacesafterend}

\def\ger{\mathfrak}

\newcommand\CategoryTypeface{\mathbf}
\def\cat{\CategoryTypeface}

\newcommand\SheafTypeface{\mathcal}
\def\sh{\SheafTypeface}

\def\DMO{\DeclareMathOperator}

\newcommand\ev{{\bar 0}}
\newcommand\odd{{\bar 1}}

\newcommand{\defi}{\coloneqq}     

\def\diff#1^#2{\ensuremath{\partial_{#1}^{#2}}}

\def\der#1/#2{\ifthenelse {\equal{#1}{}}
              {\ensuremath{\partial_{#2}{#1}}}
              {\ensuremath{\frac{\partial #1}{\partial #2}}}
        }

\def\derf#1/#2{\ifthenelse  {\equal{#1}{}}
              {\ensuremath{\frac{\partial #1}{\partial #2}}}
              {\ensuremath{\partial_{#2}{#1}}}
        }

\newcommand\cf{\emph{cf.}~}

\newcommand\ie{\emph{i.e.}~}
\newcommand\viz{\emph{viz.}~}
\newcommand\via{\emph{via}~}

\def\multi(#1,#2){\ifthenelse {\equal{#1}{0}}
                {{\mathbb Z}_2^{#2}}
                {\ifthenelse{\equal{#2}{0}}
                      {{\mathbb N}_0^{#1}}
                      {\ensuremath{{{\mathbb N}_0^{#1}\!\times\!{\mathbb Z}_2^{#2}}}}
                }
        }

\newcommand\vphi{\varphi}

\newcommand\eps{\varepsilon}

\newcommand\sle{\leqslant}

\DMO\dom{\mathrm{dom}}
\DMO\rk{\mathrm{rk}}
\DMO\Ad{\mathrm{Ad}}
\DMO\ad{\mathrm{ad}}
\DMO\GL{\mathrm{GL}}
\DMO\id{\mathrm{id}}
\DMO\pr{\mathrm{pr}}
\DMO\gr{\mathrm{gr}}
\DMO\sll{\ger{sl}}
\DMO\sdim{\mathrm{sdim}}
\DMO\sgn{\mathrm{sgn}}
\DMO\re{\mathrm{Re}}
\DMO\Gal{\mathrm{Gal}}
\DMO\Ann{\mathrm{Ann}}
\DMO\coker{\mathrm{coker}}
\DMO\im{\mathrm{im}}
\DMO\coim{\mathrm{coim}}
\DMO\Spec{\mathrm{Spec}}
\DMO\codim{\mathrm{codim}}
\DMO\chr{\mathrm{char}}
\DMO\supp{\mathrm{supp}}
\DMO\str{\mathrm{str}}
\DMO\tr{\mathrm{tr}}

\DMO\Top{\cat{Top}}
\DMO\Sets{\cat{Sets}}
\DMO\SMan{\cat{SMan}}
\DMO\SRSp{\cat{SRSp}}
\DMO\SSp{\cat{SSp}}
\DMO\SVSp{\cat{SVSp}}
\DMO\SVec{\cat{SVec}}
\DMO\SAlg{\cat{SAlg}}
\DMO\Mod{\cat{Mod}}
\DMO\Op{\cat{Op}}
\DMO\Cov{\cat{Cov}}

\DMO\Ob{\mathrm{Ob}}

\newcommand{\lBr}{[\kern-.65ex[}
\newcommand{\rBr}{]\kern-.65ex]}

\makeatletter
\newcommand\Size[7][1]{
                                 \ifx#20%
                                        \def\r@l{}\def\r@m{}\def\r@r{}%
                                 \else%
                                    \ifx#21%
                                           \def\r@l{\bigl}\def\r@r{\bigr}\def\r@m{\bigm}%
                                    \else%
                                           \ifx#22%
                                                 \def\r@l{\Bigl}\def\r@r{\Bigr}\def\r@m{\Bigm}%
                                            \else%
                                                 \ifx#23%
                                                        \def\r@l{\biggl}\def\r@r{\biggr}\def\r@m{\biggm}%
                                                  \else
                                                        \ifx#24%
                                                        \def\r@l{\Biggl}\def\r@r{\Biggr}\def\r@m{\Biggm}%
                                                        \fi%
                                                  \fi%
                                            \fi%
                                      \fi%
                                 \fi%
                                 \ifx#10%
                                       \def\r@m{}%
                                 \fi%
                                 \r@l#3{#4}\r@m#5{#6}\r@r#7%
}%

\makeatother

\makeatletter
\def\Set@Scallop[#1]#2#3{{#1}\Parens{#2}{#3}}
\newcommand\DeclareScalableOperator[2]{%
  \expandafter\def\csname#1\endcsname{\@ifnextchar[{{#2}\Set@Scallop}{{#2}\Set@Scallop[{}]}}
}
\makeatother

\def\DSO{\DeclareScalableOperator}

\DSO{Presh}{\cat{Presh}}
\DSO{Sh}{\cat{Sh}}

\DSO{ShHom}{\sh Hom} 
\DSO{ShGHom}{\underline{\sh Hom}} 
\DSO{ShEnd}{\sh End} 
\DSO{ShGEnd}{\underline{\sh End}} 
\DSO{ShDer}{\sh Der} 
\DSO{ShGDer}{\underline{\sh Der}} 
\DSO{ShCliff}{\sh C\text{\emph{liff}}} 

\DSO{Ct}{\mathcal{C}} 
\DSO{Cc}{\mathcal{C}_c} 
\DSO{Lp}{\mathbf{L}} 
\DSO{Hom}{\mathrm{Hom}} 
\DSO{End}{\mathrm{End}} 
\DSO{Aut}{\mathrm{Aut}} 
\DSO{Uenv}{\mathfrak{U}} 
\DSO{Cuenv}{\widehat{\mathfrak{U}}} 
\DSO{ABer}{\Abs0{\mathrm{Ber}}}
\DSO{Proj}{\mathrm{Proj}}
\DSO{Ber}{\mathrm{Ber}}
\DSO{Vol}{\mathrm{Vol}}
\DSO{Form}{\Omega}
\DSO{Ind}{\mathrm{Ind}}
\DSO{Coind}{\mathrm{Coind}}
\DSO{Der}{\mathrm{Der}}
\DSO{GDer}{\underline{\mathrm{Der}}}
\DSO{Alg}{\mathrm{Alg}}
\DSO{GHom}{\underline{\mathrm{Hom}}} 
\DSO{GEnd}{\underline{\mathrm{End}}} 
\DSO{Maps}{\mathrm{Maps}}
\DSO{Cliff}{\mathrm{Cliff}} 
\DSO{Jac}{\mathrm{Jac}} 

\newcommand\Set[3]{
                                 \Size{#1}{\{}{#2}{|}{#3}{\}}%
}%
\newcommand\Dual[3]{
                                 \Size[0]{#1}{\langle}{#2}{,}{#3}{\rangle}%
}%
\newcommand\Parens[2]{
  \Size[0]{#1}{(}{#2}{}{}{)}
}

\newcommand\Norm[2]{
  \Size[0]{#1}{\lVert}{#2}{}{}{\rVert}
}
\newcommand\Abs[2]{
  \Size[0]{#1}{\lvert}{#2}{}{}{\rvert}
}

\makeatletter

\makeatletter

\newif\if@smallmat
\newif\if@none
\newif\if@paren
\newif\if@brack
\newif\if@brace
\newif\if@vline
                                 {\ifx#20%
                                        \@smallmattrue%
                                  \else%
                                         \@smallmatfalse
                                  \fi%
                                  \ifx#11%
                                         \@nonefalse\@parentrue\@brackfalse\@bracefalse\@vlinefalse%
                                  \else%
                                       \ifx#12%
                                            \@nonefalse\@parenfalse\@bracktrue\@bracefalse\@vlinefalse%
                                        \else%
                                            \ifx#13%
                                                 \@nonefalse\@parenfalse\@brackfalse\@bracetrue\@vlinefalse%
                                            \else%
                                                 \ifx#14%
                                                       \@nonefalse\@parenfalse\@brackfalse\@bracefalse\@vlinetrue
                                                 \else%
                                                       \ifx#15%
                                                             \@nonefalse\@parenfalse\@brackfalse\@bracefalse\@vlinefalse%
                                                       \else%
                                                             \@nonetrue\@parenfalse\@brackfalse\@bracefalse\@vlinefalse%
                                                       \fi%
                                                 \fi%
                                            \fi%
                                        \fi%
                                   \fi%
                                   \if@smallmat%
                                        \if@none%
                                             \begin{smallmatrix}%
                                        \else%
                                            \if@paren%
                                                  \left(\begin{smallmatrix}%
                                            \else%
                                                  \if@brack%
                                                          \left[\begin{smallmatrix}%
                                                  \else%
                                                          \if@brace%
                                                               \left\{\begin{smallmatrix}%
                                                          \else%
                                                               \if@vline%
                                                                    \left\lvert\begin{smallmatrix}%
                                                                \else%
                                                                    \left\lVert\begin{smallmatrix}%
                                                                \fi%
                                                          \fi%
                                                  \fi%
                                            \fi%
                                        \fi%
                                   \else%
                                        \if@none%
                                             \begin{matrix}%
                                        \else%
                                            \if@paren%
                                                  \begin{pmatrix}%
                                            \else%
                                                  \if@brack%
                                                          \begin{bmatrix}%
                                                  \else%
                                                          \if@brace%
                                                               \begin{Bmatrix}%
                                                          \else%
                                                               \if@vline%
                                                                    \begin{vmatrix}%
                                                                \else%
                                                                    \begin{Vmatrix}%
                                                                \fi%
                                                          \fi%
                                                  \fi%
                                            \fi%
                                        \fi%
                                   \fi}%
                                  {\if@smallmat%
                                        \if@none%
                                             \end{smallmatrix}%
                                        \else%
                                            \if@paren%
                                                  \end{smallmatrix}\right)%
                                            \else%
                                                  \if@brack%
                                                          \end{smallmatrix}\right]%
                                                  \else%
                                                          \if@brace%
                                                               \end{smallmatrix}\right\}%
                                                          \else%
                                                               \if@vline%
                                                                    \end{smallmatrix}\right\rvert%
                                                                \else%
                                                                    \end{smallmatrix}\right\rVert%
                                                                \fi%
                                                          \fi%
                                                  \fi%
                                            \fi%
                                         \fi%
                                   \else%
                                        \if@none%
                                             \end{matrix}%
                                        \else%
                                            \if@paren%
                                                  \end{pmatrix}%
                                            \else%
                                                  \if@brack%
                                                          \end{bmatrix}%
                                                  \else%
                                                          \if@brace%
                                                               \end{Bmatrix}%
                                                          \else%
                                                               \if@vline%
                                                                    \end{vmatrix}%
                                                                \else%
                                                                    \end{Vmatrix}%
                                                                \fi%
                                                          \fi%
                                                  \fi%
                                            \fi%
                                        \fi%
                                   \fi}%
                         
\makeatother

\def\clap#1{\hbox to 0pt{\hss#1\hss}} 

\newcommand\C{\mathbb{C}}
\DMO\Mp{\mathrm{Mp}}
\newcommand\N{\mathbb{N}}
\newcommand\R{\mathbb{R}}
\DMO\Sp{\mathrm{Sp}}
\DMO\SO{\mathrm{SO}}
\DMO\Spin{\mathrm{Sp}}
\DMO\Herm{\mathrm{Herm}}
\newcommand\Z{\mathbb{Z}}
\DSO{WC}{\mathrm{WCl}}
\DMO\LT{\mathscr L}
\DeclareScalableOperator{Sw}{\mathscr{S}}
\DeclareScalableOperator{TDi}{\mathscr{S}'}
\DMO\Gr{\mathrm{Gr}}
\DMO\diag{\mathrm{diag}}

\begin{document}

\title[Two proofs of the superbosonisation identity]{Superbosonisation, Riesz superdistributions, and highest weight modules (extended version)}

\author[Alldridge]
{Alexander Alldridge}

\address{Universit\"at zu K\"oln\\
Mathematisches Institut\\
Weyertal 86-90\\
50931 K\"oln\\
Germany}
\email{alldridg@math.uni-koeln.de}

\author[Shaikh]{Zain Shaikh}

\address{Universit\"at zu K\"oln\\
Mathematisches Institut\\
Weyertal 86-90\\
50931 K\"oln\\
Germany}
\email{zain@math.uni-koeln.de}

\begin{abstract}
  This is the extended version of a survey prepared for publication in the Springer INdAM series. 

  Superbosonisation, introduced by Littelmann--Sommers--Zirnbauer, is a generalisation of bosonisation, with applications in Random Matrix Theory and Condensed Matter Physics. We link the superbosonisation identity to Representation Theory and Harmonic Analysis and explain two new proofs, one \via the Laplace transform and one based on a multiplicity freeness statement.
\end{abstract}

\maketitle

\section{Introduction}
\label{sec:1}

Supersymmetry (SUSY) has its origins in Quantum Field Theory. It is usually associated with High Energy Physics, especially with SUGRA, where the fermionic fields correspond to physical quantities, the mathematical incarnation of a (as yet, hypothetical) fundamental phenomenon. However, beyond this fascinating and deep theory, and its independent mathematical interest, SUSY also has applications in quite different areas of physics, notably, in Condensed Matter. 

Here, the generators of supersymmetry do not correspond to physical quantities. Rather, they appear as effective symmetries of models for low-temperature limits of the fundamental Quantum Field Theory. This idea goes under the name of the \emph{Supersymmetry Method}, and was developed, by Efetov and Wegner \cite{efetov-susy}. 

Its particular merit is the possibility to derive, by the use of Harmonic Analysis on certain symmetric superspaces, precise closed form expressions for statistical quantities---such as the moments of the conductance of a metal with impurities \cite{zirn-cmp,zirn-prl}---in a regime where the system becomes critical, for instance, exhibiting a transition from localisation to diffusion, which is not tractable by other methods. 

In connection with the physics of thin wires, the subject was well studied in the 1990s; it has recently gained substantial new interest, since the `symmetry classes' investigated in this context \cite{zirn-rmt,az-10fold,hhz} have been found to occur as `edge modes' of certain 2D systems dubbed `topological insulators' (resp.~superconductors) \cite{freed-moore}.

Mathematically, several aspects of the method beg justification. One both subtle and salient point is the transformation of certain integrals over flat superspace in high dimension $N\to\infty$, which occur as expressions for statistical Green's functions, into integrals over a curved superspace of fixed rank and dimension---the latter being more amenable to asymptotic analysis (by steepest descent or stationary phase). Traditionally, this step is performed by the use of the so-called Hubbard--Stratonovich transformation, which is based on a careful deformation of the integration contour. 

This poses severe analytical problems, which to the present day have only been overcome in cases derived from random matrix ensembles that follow the normal distribution \cite{hill-zirn}. To extend the Supersymmetry Method's range beyond Gaussian disorder, for instance to establish universality for invariant random matrix ensembles, a complementary tool was introduced, based on ideas of Fyodorov \cite{fyodorov}: the \emph{Superbosonisation Identity} proved by Littelmann--Sommers--Zirnbauer in their seminal paper \cite{lsz}.

\bigskip\noindent
We now proceed to describe this identity. In general, it holds in the context of unitary, orthogonal, and unitary-symplectic symmetry. We restrict ourselves to the first case (of unitary symmetry), although our methods carry over to the other cases. 

One considers the space $W\defi\C^{p|q\times p|q}$ of square super-matrices and a certain subsupermanifold $\Omega$ of purely even codimension, whose underlying (Riemannian symmetric) manifold $\Omega_0$ is the product of the positive Hermitian $p\times p$ matrices with the unitary $q\times q$ matrices. Let $f$ be a superfunction defined and holomorphic on the tube domain based on $\Herm^+(p)\times\Herm(q)$. The superbosonisation identity states 
\begin{equation}\label{eq:sbos}
  \int_{\C^{p|q\times n}\oplus\C^{n\times p|q}}\Abs0{Dv}\,f(Q(v))=C\int_\Omega\Abs0{Dy}\,\Ber0y^nf(y),
\end{equation}
for some finite positive constant $C$, provided $f$ has sufficient decay at infinity along the manifold $\Omega_0$. Here, $Q$ is the quadratic map $Q(v)=vv^*$, $\Abs0{Dv}$ is the flat Berezinian density, and $\Abs0{Dy}$ is a Berezinian density on $\Omega$, invariant under a certain natural transitive supergroup action we will specify below. 

Remark that any $\GL(n,\C)$-invariant superfunction on $\C^{p|q\times n}\oplus\C^{n\times p|q}$ may be written in the form $f(Q(v))$. A notable feature of the formula is thus that it puts the `hidden supersymmetries' (from $\GL(p|q,\C)$) into evidence through the invariant integral over the homogeneous superspace $\Omega$, where `manifest symmetries' (from $\GL(n,\C)$) enter \via some character (namely, $\Ber0y^n$).

A remarkable special case occurs when $p=0$. Then Equation \eqref{eq:sbos} reduces to
\[
  \int_{\C^{0|q\times n}\oplus\C^{n\times 0|q}}\Abs0{Dv}\,f(Q(v))=C\int_{\mathrm U(q)}\Abs0{Dy}\,\det(y)^{-n}f(y),
\]
which is known as the \emph{Bosonisation Identity} in physics. Notice that the left-hand side is a purely fermionic Berezin integral, whereas the right-hand side is purely bosonic. Formally, it turns fermions $\psi\bar\psi$ into bosons $e^{i\vphi}$. If in addition $q=1$, we obtain the \emph{Cauchy Integral Formula}. 

At the other extreme, if $q=0$, then $\Omega=\Herm^+(p)$, and Equation \eqref{eq:sbos} is a classical identity due to Ingham and Siegel \cite{ingham,siegel}, well-known to harmonic analysts. It admits a far-reaching generalisation in the framework of Euclidean Jordan algebras \cite{FK94}. The first one to use it in the physics context that inspired superbosonisation was Fyodorov \cite{fyodorov}. Moreover, the right-hand side of the identity, \viz
\[
  \Dual0{T_n}f\defi\int_{\Herm^+(p)}\Abs0{Dy}\,\det(y)^nf(y)
\]
is the so-called (unweighted) \Define{Riesz distribution}. After suitable renormalisation, it becomes analytic in the parameter $n$, a fact that was exploited in the analytic continuation of holomorphic discrete series representations by Rossi--Vergne \cite{rv}. See also Ref.~\cite{fk}.

\medskip\noindent
This observation links the identity to equivariant geometry and Lie theoretic Representation Theory, and this was our motivation to re-investigate the identity.

In this survey, we explain two new proofs of the superbosonisation identity, which exploit these newly found connections. One of these proofs is based on Representation Theory. Namely, as it turns out, the two sides of the identity are given by certain special relatively invariant functionals on two highest weight modules of the Lie superalgebra $\ger g'\defi\ger{gl}(2p|2q,\C)$, which are infinite dimensional for $p>0$. Their equality is an immediate consequence of a multiplicity one statement. We will explain the main ingredients of the proof; details shall be published elsewhere. 

On the other hand, to actually identify the r.h.s.~as a functional on the corresponding representation requires the construction of an intertwining operator in the form of a certain weighted Laplace transform $\LT_n$. This leads to another proof, based on comparing Laplace transforms. The functional analytic details of this proof can be found in Ref.~\cite{as-sbos}. Here, we only explain the main ideas.

Combining both points leads to further developments. Indeed, the representations related to the superbosonisation identity depend on a parameter $n\geqslant p$. Using functional equations, one may show that the r.h.s.~is analytic as a distribution-valued function of $n$. In a forthcoming paper, we shall use this fact to investigate the analytic continuation of the representations.

\bigskip\noindent\textit{Acknowlegment.} This research was funded by the grants no.~DFG ZI 513/2-1 and SFB TR/12, provided by Deutsche Forschungsgemeinschaft (DFG). We wish to thank Martin Zirnbauer for extensive discussions, detailed comments, and for bringing this topic to our attention. We thank INdAM for its hospitality. The first named author wishes to thank Jacques Faraut for his interest, and the second named author wishes to thank Bent \O{}rsted for some useful comments. 

\medskip\noindent
In what follows, we use the terminology of supergeometry freely. For the reader's convenience, some basics are summarised in an Appendix. 

\section{The superbosonisation module}
\label{sec:2}
In this section, we define, by the use of Howe duality, a particular supermodule, which will turn out to be intimately related to the superbosonisation identity.

\subsection{The oscillator representation}

We begin by reviewing some standard material on the Weyl--Clifford algebra in a form well suited for our purposes. Consider $V\defi U\oplus U'$ where 
\[
	U\defi\GHom0{\C^n,\C^{p|q}},\quad U'\defi\GHom0{\C^{p|q},\C^n},
\]
and $\GHom0{\cdot,\cdot}$ denotes the set of all linear maps with its usual $\Z/2\Z$ grading. Then $V^*=U^*\oplus U^{\prime *}$. The natural supersymplectic form on $V\oplus V^*$ is
\[
	s(a+f,b+g)\defi f(b)-(-1)^{|a||g|}g(a),
\]
where $a,b\in V$ and $f,g \in V^*$ are homogeneous.

The trace form on $\GEnd0{\C^n}$ and the supertrace form on $\GEnd0{\C^{p|q}}$, respectively, allow us to identify
\[
	V^*=\GHom0{\C^{p|q},\C^n}\oplus\GHom0{\C^n,\C^{p|q}}.
\]
In these terms, $s((u,\varphi,\gamma,x),(v,\psi,\delta,y))$ can be rewritten as
\begin{equation}\label{eqn:form}
  \tr(\gamma v)+\str(x \psi)-(-1)^{|u||\delta|}\tr(\delta u)-(-1)^{\Abs0y\Abs0\vphi}\str(y \varphi),
\end{equation}
for homogeneous $(u,\varphi),(v,\psi)\in V$ and $(\gamma,x),(\delta,y)\in V^*$.

Let $\ger h_V$ be the central extension of the Abelian Lie superalgebra $V\oplus V^*$ by $\C\mathbf1$ that is determined by $s$. Its non-zero bracket relations are 
\[ 
	[v,w]=s(v,w)\mathbf{1}
\]
for $v,w\in V\oplus V^*$. By definition, the \Define{Weyl--Clifford algebra} is 
\[
	\WC0V\defi\Uenv0{\ger h_V}/(\mathbf{1}-1),
\] 
where $\mathbf1\in\ger h_V\subseteq\Uenv0{\ger h_V}$ and $1\in\Uenv0{\ger h_V}$ is the unit of the universal enveloping algebra. 

\subsubsection*{Canonical $\ger{spo}$ subalgebra}

The Weyl--Clifford algebra inherits an ascending filtration $\WC[_n]0V$ from the tensor algebra $\bigotimes(V\oplus V^*)$. The PBW theorem implies that $\gr\WC0V=S(V\oplus V^*)$, the supersymmetric algebra of $V\oplus V^*$. 

Conversely, $\WC0V$ inherits a canonical augmentation from $\Uenv0{\ger h_V}$; moreover, the kernel of the canonical map $\bigotimes(V\oplus V^*)\to\WC0V$ is generated by quadratic relations without linear term, and it follows that there is a canonical splitting of the map $\WC0V\to S(V\oplus V^*)$ in degree two, the image of which we denote by $\ger s$.

Explicitly, for $a,b\in V\oplus V^*$, the quadratic element $ab\in S^2(V\oplus V^*)$ is embedded into $\WC0V$ as
\[
	\textstyle{\frac12}\Parens1{ab+(-1)^{|a||b|}ba}.
\]
On general grounds, the bracket satisfies $[\WC[_a]0V,\WC[_b]0V]\subseteq\WC[_{a+b-1}]0V$, so $\ad\ger s$ preserves the filtration; since, moreover, $\C1$ is central in $\WC0V$, $\ger s$ is a Lie superalgebra and the degree one part $\gr_1\WC0V\cong V\oplus V^*\subseteq\WC0V$ is an $\ger s$-module. In fact, this sets up an isomorphism $\ger s\cong\ger{spo}(V\oplus V^*)$, where the latter is the Lie subsuperalgebra of $\ger{gl}(V\oplus V^*)$ consisting of those endomorphisms that leave the form $s$ infinitesimally invariant.

\subsubsection*{Oscillator module}

Let $V\oplus V^*=X\oplus Y$ be a complex polarisation of $V\oplus V^*$. That is, $X$ and $Y$ are maximal isotropic subspaces. The Weyl--Clifford algebra $\WC0V$ has a module
\[
	S_X\defi\WC0V/\WC0V\cdot X=\Uenv0{\ger h_V}\otimes_{\Uenv0{X\oplus\C\mathbf1}}\C.
\]
called the \Define{oscillator} (or \emph{spinor}) \emph{representation}. Here, $\C$ is understood to be the module of the Abelian Lie subsuperalgebra $V\oplus\C1\subseteq\ger h_V$ on which $\mathbf1$ acts as the identity and $X$ acts by zero.

We have $S_X\cong S(Y)=\C[X]$ as modules over the Abelian Lie subsuperalgebra $Y$ of $\ger h_V$, where the action of $y\in Y$ on $f\in\C[X]$ is defined by left multiplication:
\[
	y\cdot f:=\ell_yf\defi yf.
\]
Here, we consider $Y\subseteq\C[X]$ by defining $y(x)\defi s(y,x)$ for all $y\in Y$, $x\in X$. By our definition of $s$, this recovers the usual meaning of $v^*(v)$ for $X=V$, $Y=V^*$.

In these terms, $\mathbf1$ acts as the identity, and $x\in X$ acts by the unique superderivation of the algebra $\C[X]$ that is defined on degree one elements $f\in Y\subseteq\C[X]$ by 
\[
	x\cdot f:=\partial_xf\defi s(x,f).
\]
Since any $f\in\C[X]$ that is annihilated by all $\partial_x$ is necessarily constant, the maximal proper submodule of $S_X$ is zero, and $S_X$ is a simple module over $\WC0V$. Since $\ger s$ is a Lie subsuperalgebra of $\WC0V$, $S_X$ is also an $\ger s$-module.

The isomorphism class of $S_X$ as a $\WC0V$-module does not depend on the choice of the polarisation. For the particular choice $X\defi V$, $Y\defi V^*$, we have $S_X=\C[V]$. However, we will need to allow for other choices of polarisation, in to order to accommodate the action of real forms of $\ger s$ or its subalgebras. 

\subsection{An application of Howe duality}

Let $G_\C\defi\GL(n,\C)$ be the complexification of the unitary group $G=\mathrm U(n)$. Define an action of $G_\C$ on $V$ by
\[
	g\cdot (u,\varphi)\defi(ug^{-1},g\varphi),
\]
for $g\in G_\C$, $(u,\varphi)\in V$; on $V^*$, we have the contragredient $G_\C$-action. The induced action of $G_\C$ on $V\oplus V^*$ is faithful and preserves the form $s$; hence, it realises the Lie algebra $\ger g$ of $G_\C$ as a subalgebra of $\ger s$.

Let $\ger g'\defi\ger z_{\ger s}(\ger g)$ be the centraliser of $\ger g$ in $\ger s$. Then $\ger g'\cong\ger{gl}(2p|2q,\C)$ and $\ger z_\ger s(\ger g')=\ger g$. In other words, $(\ger g,\ger g')$ form a \emph{dual pair}, see Ref.~\cite{howe-classical}. This is a special case of the following simple lemma (\cf Ref.~\cite{howe-classical}), which we apply with $A=\C^n$, $B=\C^{2p|2q}$ and the isomorphism
\[
\Psi:V\oplus V^*\to\GHom0{A,B}\oplus\GHom0{B,A}:(u,\varphi,\gamma, x)\longmapsto
\left(\begin{pmatrix}
u\\
-x\end{pmatrix}, (\gamma\ \varphi)\right).
\]
of super-symplectic vector spaces.

\begin{Lem}
	Let $A,B$ be super-vector spaces and $\ger{spo}\defi\ger{spo}(\GHom0{A,B}\oplus\GHom0{B,A})$. Then we have 
	\[
		\ger z_{\ger{spo}}(\ger{gl}(A))=\ger{gl}(B)\ ,\ \ger z_{\ger{spo}}(\ger{gl}(B))=\ger{gl}(A).
	\] 
	Here, the embedding of $\ger{gl}(C)$ into $\ger{spo}$ (where $C=A,B$) is given by the natural action of the former on $\GHom0{A,B}\oplus\GHom0{B,A}$.
\end{Lem}

Explicitly, the action of $\ger g'=\ger{gl}(2p|2q,\C)$ on $V\oplus V^*$---determining its embedding into $\ger s$---is given as follows. For $Y\in \ger{g}^\prime$ decomposed as
\begin{equation}\label{eqn:gprime}
  \kbordermatrix{~ & p|q & p|q \cr
  p|q & A & B \cr
  p|q & C & D},
\end{equation}
the action $Y$ of $V\oplus V^*$ is induced by the isomorphism $\Psi$ given by the matrix 
\begin{equation}\label{eqn:gprimespo}
\kbordermatrix{~ & U & U^\prime &  & U^* & U^{\prime *} \cr
U & L_A & 0 & \vrule & 0 & -L_B \cr
U^\prime & 0 & -R_D & \vrule & -R_B & 0 \cr
\hline
U^* & 0 & -R_C & \vrule & -R_A & 0 \cr
U^{\prime *} & -L_C & 0 & \vrule & 0 & L_D}
\end{equation}
where $L$ and $R$, respectively, denote left and right multiplication. 

Howe's celebrated duality \cite[Theorem 8]{howe-classical}, leads to the following conclusion: 

\begin{Prop}[simple]
  The $\ger g'$-submodule $\C[V]^{G_\C}=\C[V]^G$ of $\C[V]$ formed by the $G_\C$-invariant superpolynomials is simple.
\end{Prop}

\subsubsection*{Short $\Z$-gradings}

In order to identify the representation considered in \thmref{Prop}{simple} in terms of highest weights, we introduce a triangular decomposition of $\ger s=S^2(V\oplus V^*)$. Define
\[
  \ger s_+\defi S^2(V),\quad \ger s_0\defi V\otimes V^*,\quad \ger s_-\defi S^2(V^*).
\]
This defines a $\Z$-grading of $\ger s$, \ie $\ger s_\pm$ are Abelian Lie subsuperalgebras, normalised by the Lie subsuperalgebra $\ger s_0$, and $[\ger s_+,\ger s_-]\subseteq\ger s_0$.

Indeed, $\ger s_0$ is the subalgebra of $\ger s$ that leaves the summands $V$ and $V^*$ of $V\oplus V^*$ invariant. Similarly, $\ger s_+$ (resp.~$\ger s_-$) annihilates $V$ (resp.~$V^*$) and maps $V^*$ to $V$ (resp.~$V^*$ to V). The triangular decomposition can be written in matrix form as:
\begin{equation}\label{eqn:spodecomp}
  \kbordermatrix{~ & V & V^* \cr
  V & 0 & + \cr
  V^* & - & 0}.
\end{equation}

We observe that $\ger g\subseteq\ger s_0$, so that it preserves the decomposition. Hence, $\ger g'$ inherits a triangular decomposition
\[
  \ger p^+\defi\ger g'_+\defi\ger g'\cap\ger s_+,\quad \ger k\defi\ger g'_0\defi\ger g'\cap\ger s_0,\quad \ger p^-\defi\ger g'_-\defi\ger g'\cap\ger s_-
\]
from $\ger s$. Comparing Equations \eqref{eqn:gprimespo} and \eqref{eqn:spodecomp}, for $Y$ in the form \eqref{eqn:gprime}, we have $(A,D)\in \ger k$, $B\in \ger g'_+$ and $C\in \ger g'_-$, \ie
\begin{equation}\label{eqn:gprimegrad}
  \ger g'=\kbordermatrix{~ & p|q & p|q \cr
  p|q & \ger k_1 & \ger p_+ \cr
  p|q & \ger p_- & \ger k_2},
\end{equation}
where $\ger k_j$, $j=1,2$, denote the two copies of $\ger{gl}(p|q,\C)$ in $\ger k$. Both of $\ger p_\pm=\ger g'_\pm$ are the Abelian Lie superalgebra isomorphic to $\ger{gl}(p|q,\C)$ as a super-vector space. The subalgebra $\ger k$ can also be characterised as largest subalgebra of $\ger g'=\ger{gl}(2p|2q,\C)$ that preserves the decomposition of $V$ as $U\oplus U^\prime$, \cf Ref.~\cite{howe-classical}.

\subsubsection*{Highest weight for the Borel subalgebra adapted to the grading}

From the definitions, we see that the constants $\C1\subseteq\C[V]$ are annihilated by $\ger s_+$ and left invariant as a subspace by the action of $\ger s_0$. We have the following result.

\begin{Prop}
  The simple $\ger g'$-module $\C[V]^{G_\C}$ has a highest weight. For a Borel subsuperalgebra $\ger b\subseteq\ger g'$ contained in $\ger k\oplus\ger p^+$, it is given as the restriction to a Cartan subalgebra of the $\ger k$-character $\lambda$ defined by the formula
  \[
    \lambda\begin{pmatrix}
    A & 0\\
    0 & D
    \end{pmatrix}=\frac{n}{2}(\str(D)-\str(A)).
  \]
\end{Prop}

Since $\ger g'$ inherits the grading, it suffices to compute the $\ger k$-action. To that end, we realise $\ger g'$ inside $\WC0V$. By definition, 
\begin{align*}
U&= \C^{p|q}\otimes(\C^n)^*,\quad U^\prime= \C^{n}\otimes(\C^{p|q})^*, \\
 U^*&= \C^{n}\otimes(\C^{p|q})^*, \quad U^{\prime *}=\C^{p|q}\otimes(\C^n)^*,
\end{align*}
so define two copies of homogeneous bases $(x_a)$, $(y_a)$ for the copies of $\C^{p|q}$ in $U$ and $U^{\prime *}$ (resp.), with dual bases $(x^a)$, $(y^a)$. Also, let $(e_b)$ and $(e^b)$ be dual bases of $\C^n$ and $(\C^n)^*$ respectively. A basis for $\ger g'$ is given by 
\[
E^A_{ij}=x_ix^j, \quad E^B_{ij}=x_iy^j,\quad E^C_{ij}=y_ix^j, \quad E^D_{ij}=y_iy^j,
\]
following the decomposition in Equation \eqref{eqn:gprime}. One can see that $E_{ij}^A$ is mapped to
\[
  -\sum_{\alpha=1}^n (x_i e^\alpha)\cdot(e_\alpha  x^j)\in\WC0V,
\]
which acts in the oscillator representation by the operator
\[
  T_{E^A_{ij}}:=-\frac12\sum_{\alpha=1}^n (\partial_{x_i e^\alpha}\ell_{e_\alpha x^j}+(-1)^{|i||j|}\ell_{e_\alpha x^j}\partial_{x_i e^\alpha}).
\]
Hence, $\lambda(E^A_{ij})$ is given by
\[
T_{E^A_{ij}}1= -\frac12\sum_{\alpha=1}^n(-1)^{|i||j|} \langle e_\alpha x^j, x_i  e^\alpha \rangle= -\frac{n}{2}(-1)^{|i||j|}\delta_{ij}= -\frac{n}{2}\str E^A_{ij},
\]
where $|i|$ is the parity of $x_i$ and $\delta_{ij}$ is the Kronecker delta. The calculation of $\lambda(E^D_{ij})=\frac{n}{2}\str E^D_{ij}$ is completely analogous.

\subsubsection*{Highest weight for the standard Borel}

In order to determine the highest weight for the standard Borel subsuperalgebra, let us be more explicit. We take $\ger h$ to be the Cartan subalgebra defined as the span of $E^A_{aa}$, $1\sle a\sle p+q$ and $E^D_{dd}$, $1\sle d\sle p+q$. A basis $\delta_i,\eps_j$, $1\sle i\sle 2p$, $1\sle j\sle 2q$, of $\ger h^*$ is given by 
\begin{align*}
  \delta_i(E^A_{aa})&\defi\delta_{ia}\theta(p-a)\theta(p-i),\\
  \delta_i(E^D_{dd})&\defi\delta_{i-p,d}\theta(p-d)\theta(i-p-1),\\
  \eps_j(E^A_{aa})&\defi\delta_{j,a-p}\theta(a-p-1)\theta(q-j),\\
  \eps_j(E^D_{dd})&\defi\delta_{j-q,d-p}\theta(d-p-1)\theta(j-q-1),
\end{align*}
where we agree to write
\[
  \theta(x)\defi
  \begin{cases}
    1 & x\geqslant0,\\
    0& x<0.
  \end{cases}
\]

Let $\ger b\subseteq\ger g'$ be the Borel subsuperalgebra determined uniquely by $\ger b\cap\ger k$ being the direct product of the standard Borels for $\ger k_j$, $j=1,2$, and $\ger b\cap(\ger p^+\oplus\ger p^-)=\ger p^+$. Then $\ger b$ is contained in the parabolic subalgebra $\ger k\oplus\ger p^+$, and its Dynkin diagram is:

\smallskip
\begin{center}
  \begin{tikzpicture}
    \node[root, label=below:$\delta_1-\delta_2$] (a) {}; 
    \node[root, label=above:$\delta_{p-1}-\delta_p$] (b) [right=2.2em of a] {}; 
    \node[io-root, label=below:$\delta_p-\eps_1$] (c) [right=2.2em of b] {}; 
    \node[root, label=above:$\eps_1-\eps_2$] (d) [right=2.2em of c] {}; 
    \node[root, label=below:$\eps_{q-1}-\eps_q$] (e) [right=2.2em of d] {}; 
    \node[io-root, label=above:$\eps_q-\delta_{p+1}$] (f) [right=2.2em of e] {}; 
    \node[root, label=below:$\delta_{p+1}-\delta_{p+2}$] (g) [right=2.2em of f] {}; 
    \node[root, label=above:$\delta_{2p-1}-\delta_{2p}$] (h) [right=2.2em of g] {}; 
    \node[io-root, label=below:$\delta_{2p}-\eps_{q+1}$] (i) [right=2.2em of h] {}; 
    \node[root, label=above:$\eps_{q+1}-\eps_{q+2}$] (j) [right=2.2em of i] {}; 
    \node[root, label=below:$\eps_{2q-1}-\eps_{2q}$] (k) [right=2.2em of j] {}; 
    \draw[dashed] (a)--(b);
    \draw[thick] (b)--(c);
    \draw[thick] (c)--(d);
    \draw[dashed] (d)--(e);
    \draw[thick] (e)--(f);
    \draw[thick] (f)--(g);
    \draw[dashed] (g)--(h);
    \draw[thick] (h)--(i);
    \draw[thick] (i)--(j);
    \draw[dashed] (j)--(k);
  \end{tikzpicture}
\end{center}
It is obtained from the standard Borel $\ger b_{\mathrm{st}}$ by the application of $R_{\max(p,q)}\cdots R_1$, where $R_i$ is the chain of odd reflections given by 
\[
  R_i\defi r_{\delta_{\max(p+1,2p-i+1)}-\eps_{\min(i,q)}}\cdot
  \begin{cases}
    R_i^\delta R_i^\eps & i\sle\min(p,q),\\
    R_i^\delta & q<i\sle p,\\
    R_i^\eps & p<i\sle q,
  \end{cases}
\]
where 
\[
  R_i^\delta\defi\prod_{k=1}^ir_{\delta_{2p-i+1}-\eps_{\min(k,q)}},\quad
  R_i^\eps\defi\prod_{k=1}^i r_{\delta_{\max(p+1,2p-k+1)}-\eps_i}.
\]
These products are ordered such that factors corresponding to smaller values of $k$ occur farther to the right. We have 
\[
  \lambda|_\ger h=-\frac n2\sum_{i=1}^p\delta_i+\frac n2\sum_{i=p+1}^{2p}\delta_i+\frac n2\sum_{j=1}^q\eps_j-\frac n2\sum_{j=q+1}^{2q}\eps_j,
\]
so $\Dual0\lambda{\delta_i-\eps_j}=0$ for $i>p$ and $j\sle q$. Hence, $\lambda|_{\ger h}$ is also the highest weight of $\C[V]^{G_\C}$ with respect to the standard Borel $\ger b_{\mathrm{st}}$ \cite{cw-book}, and $\C[V]^{G_\C}=L(\lambda)$. Moreover, 
\[
  \Dual0\lambda{\delta_i-\delta_{i+1}}=-\delta_{ip}n,\quad
  \Dual0\lambda{\eps_j-\eps_{j+1}}=\delta_{jq}n.
\]
By standard facts \cite{cw-book}, this proves the following statement.

\begin{Prop}
  The simple $\ger g'$-module $L(\lambda)=\C[V]^{G_\C}$ of highest weight $\lambda$ has finite dimension if and only if $p=0$. 
\end{Prop}

Note that the highest weight $\lambda|_\ger h$ is integral or half-integral, depending on whether $n$ is even or odd. Moreover, $L(\lambda)$ is atypical whenever $pq>0$.

\section{A relatively invariant functional}

In this section, we show how to realise the left hand side of the superbosonisation identity as a relatively invariant functional on a globalisation of $L(\lambda)$. 

\subsection{Globalisation of the oscillator representation}

In order to globalise the oscillator representation $\C[V]$, we need to consider real forms. We shall use the following concept, \cf Refs.~\cite{deligne-morgan,bouarroudj-grozman-leites-shchepochina}.

\begin{Def}[csvs]
  A real $\Z/2\Z$ graded vector space $U=U_\ev\oplus U_\odd$ with a fixed complex structure on $U_\odd$ will be called a \emph{\emph{cs} vector space}. Given a complex super-vector space $W$, we call a \emph{cs} vector subspace $U$ a \emph{\emph{cs} form} of $W$ if $W_\ev=U_\ev\oplus iU_\ev$ and $W_\odd=U_\odd$.
\end{Def}

We introduce a \emph{cs} form $V_\R$ of $V$ by positing 
\[
  V_{\R,\ev}\defi \mathcal X_\ev\defi V_\ev\cap\Herm(n+p)=\Set1{(L,L^*)}{L\in\C^{p\times n}},
\]
where $L^*$ denotes the conjugate transpose of the matrix $L$. Dually, we consider the \emph{cs} form of $V^*$ defined by 
\[
  V_{\R,\ev}^*\defi\mathcal Y_\ev\defi\Set1{(K^*,K)}{K\in\C^{p\times n}}.
\]
Then $s$ is real and non-degenerate on $\mathcal X_\ev\oplus\mathcal Y_\ev=V_{\R,\ev}\oplus V_{\R,\ev}^*$, and in fact, this is a totally real polarisation of $V_\ev\oplus V_\ev^*$. 

\subsubsection*{Boson-boson sector}

Let $F^0\defi L^2(\mathcal X_\ev)$, where we take the Lebesgue measure on $V_{\R,\ev}$ induced by the Euclidean form $\tr(LL^*)$. Consider the Heisenberg group $H^0$, \ie the connected and simply connected real Lie group with Lie algebra $(V_{\R,\ev}\times V_{\R,\ev}^*\times i\R)\cap\ger h_{V_\ev}$. The defining bracket relations of the latter are given by 
\[
  [(L_1,L_1^*,K_1^*,K_1),(L_2,L_2^*,K_2^*,K_2)]=2\Re\Parens1{\tr(K_1^*L_2)-\tr(K_2^*L_1)}.
\]
The Schr\"odinger model of the oscillator representation of $H^0$ on $F^0$ defines a representation of the real symplectic Lie algebra $\ger{sp}(\mathcal X_\ev\oplus\mathcal Y_\ev,\R)$, which integrates to a unitary representation of the double cover $\Mp\defi\Mp(\mathcal X_\ev\oplus\mathcal Y_\ev,\R)$ of the real symplectic group $\Sp\defi\Sp(\mathcal X_\ev\oplus\mathcal Y_\ev,\R)$ \cite{weil,nishiyama-oscillator}. 

Let $\tilde U^0$ be the lift to $\Mp$ of the maximal compact subgroup of $\Sp$. That is, we have $\tilde U^0=\mathrm U(np)\times_{\mathrm U(1)}\mathrm U(1)$, the $\sqrt{\det}$ double cover of $\mathrm U(np)$.

We consider $\widehat S_{\mathcal X_\ev}$, the formal power series ring on $\mathcal X_\ev$, and the Gaussian 
\[
  \Gamma^0\defi e^{-\tr LL^*/2}\in\widehat S_{\mathcal X_\ev}. 
\]
The action of $\ger h_{V_\ev}$ extends to this space. Then $\Gamma^0$ is annihilated by the action of $X_\ev\subseteq\ger h_{V_\ev}$, where $X_\ev\oplus Y_\ev$ is the totally complex polarisation of $V_\ev\oplus V_\ev^*$ given by 
\begin{align*}
  X_\ev&\defi\Set1{(L,K,-K,-L)}{L\in\C^{p\times n},K\in\C^{n\times p}},\\
  Y_\ev&\defi\Set1{(L,K,K,L)}{L\in\C^{p\times n},K\in\C^{n\times p}}.
\end{align*}
Thus, we have that 
\[
  \C[\mathcal X_\ev]\Gamma^0=\C[\mathcal X_\ev]e^{-\tr(LL^*)/2}\subseteq L^2(\mathcal X_\ev)=F^0
\]
is the space of $\tilde U^0$-finite vectors in $F^0$, and as an $\ger h_{V_\ev}$- and $(\ger s^0,\tilde U^0)$-module, where we define $\ger s^0\defi\ger{sp}(V_\ev\oplus V_\ev^*,\C)$, it is isomorphic to the oscillator representation $S_{X_\ev}=\C[X_\ev]$ \cite[Lemma 4.1]{adams}.

\subsubsection*{Fermion-fermion sector}

A similar argument applies to $V_\odd$, the only difference being that real forms need not to be chosen. Indeed, setting
\[
  \mathcal X_\odd\defi V_\odd,\quad\mathcal Y_\odd\defi V_\odd^*,
\]
we have a complex polarisation of $V_\odd\oplus V_\odd^*$. Thus, $F^1\defi S(V_\odd^*)=\bigwedge(V_\odd^*)=S_{\mathcal X_\odd}$ as a module of the Clifford algebra $\WC0{V_\odd}$. It contains the Gaussian
\[ 
  \Gamma^1\defi e^{\tr(K_1K_2)/2},
\]
where $K_1$ and $K_2$, respectively, denote the identity of $\C^{0|q\times n}=U^{\prime *}_\odd$ and $\C^{0|n\times q}=U^*_\odd$. Similar to the above, $\Gamma^1$ is annihilated by $X_\odd\subseteq\ger h_{V_\odd}$, where the spaces
\begin{align*}
  X_\odd&\defi\Set1{(L,K,-K,-L)}{L\in\C^{0|q\times n},K\in\C^{n\times0|q}},\\
  Y_\odd&\defi\Set1{(L,K,K,L)}{L\in\C^{0|q\times n},K\in\C^{n\times0|q}},
\end{align*}
form a complex polarisation of $V_\odd\oplus V_\odd^*$. Since $\Gamma^1$ is invertible, we have 
\[
  \C[\mathcal X_\odd]\Gamma^1=\C[\mathcal X_\odd]e^{\tr(K_1K_2)/2}=F^1,
\]
and as an $\ger h_{V_\odd}$-module, it is isomorphic to $S_{X_\odd}=\C[X_\odd]$.

\subsubsection*{Full graded picture}

Let $X\defi X_\ev\oplus X_\odd$ and observe that $\mathcal X_\ev\oplus\mathcal X_\odd=V_\R$. Then by \cite[Lemma 5.4]{nishiyama-oscillator}, we have that 
\[
  \C[V_\R]\Gamma\subseteq F\defi F^0\otimes F^1
\]
is isomorphic to $\C[X]$ as a $\ger h_V$-module, where 
\[
  \Gamma\defi\Gamma^0\cdot\Gamma^1=e^{-\str(X^2)/4},\quad X=
  \begin{pmatrix}
    0&L^*&K_2\\
    L&0&0\\
    K_1&0&0
  \end{pmatrix}.
\]

Moreover, let $\ger s^1\defi\ger{o}(V_\odd\oplus V_\odd^*,\C)$ and $\tilde U^1_\C$ the complex Lie group the Lie algebra exponentiates to in the Clifford algebra $\WC0{V_\odd}$ (\ie the the complex spin group $\Spin(nq,\C)$, the simply connected double cover of $\SO(nq,\C)$). Then $\C[V_\R]\Gamma$ is isomorphic to $\C[X]$ as $\tilde U_\C$-module, where $\tilde U_\C\defi\tilde U_\C^0\times\tilde U_\C^1$ and $\tilde U_\C^0$ is the complexification of $\tilde U^0$. That is, $\tilde U^0_\C=\GL(nq,\C)\times_{\C^\times}\C^\times$, the $\sqrt{\det}$ double cover of $\GL(nq,\C)$. In summary, $\C[V_\R]\Gamma\cong\C[X]$ as $(\ger s,\tilde U_\C)$-modules. 

\subsection{Action on Schwartz superfunctions}

Recall the terminology summarised in the Appendix. We consider $V_\R$ as a \emph{cs} manifold, namely, the \emph{cs} affine superspace associated with the \emph{cs} vector space $V_\R$.

We define $\Sw0{V_\R}\defi\Sw0{V_{\R,\ev}}\otimes\bigwedge V_\odd^*\subseteq\Gamma(\sh O_{V_\R})$, where 
\[
  \Sw0{V_{\R,\ev}}\defi\Set3{f\in\Ct[^\infty]0{V_{\R,\ev}}}{\forall p\in\C[V_{\R,\ev}]\,:\,\int_{V_{\R,\ev}}\Abs0{p(L)f(L)}^2\,dL\,d\bar L<\infty}
\]
is the Schwartz space of $V_{\R,\ev}$. We find that 
\[
  \C[V_\R]\Gamma\subseteq\Sw0V\subseteq L^2(V_{\R,0})\otimes\textstyle\bigwedge V_\odd^*=F.
\]
Since the leftmost of these is the space of $\tilde U^0\times\tilde U^1_\C$-finite vectors of the $\Mp\times U^1_\C$-module $F$, this is a chain of dense inclusions. In fact, $\Sw0{V_\R}$ is the space of smooth vectors of $F$, considered as an $\Mp\times\tilde U^1_C=\Mp(np,\R)\times\Spin(nq,\C)$-module \cite{howe-qmpde}. As one easily checks, the action of $\ger s$ extends to $\Sw0{V_\R}$, and this gives a representation of the \emph{cs} supergroup pair $(\ger s,\Mp\times\tilde U^1)$, for any real form $\tilde U^1$ of $\tilde U^1_\C$.

Let $\Abs0{Dv}$ be the Berezinian density that is associated with the standard coordinate system on $V_\R$ (see the Appendix). Then we have the following fact.

\begin{Prop}
  The Berezin integral defines a functional $\Abs0{Dv}$ on the space $\Gamma_c(\sh O_{V_\R})$ of compactly supported superfunctions by 
  \[
    \Dual0{\Abs0{Dv}}f\defi\int_{V_\R}\Abs0{Dv}\,f(v). 
  \]
  It has a unique continuous extension to $\Sw0{V_\R}$.
\end{Prop}

As $\ger s$-modules, $S_V=\C[V]$ and $S_X=\C[X]$ are isomorphic. Since both $\Gamma$ and $U,U',U^*,U'^*$ are $G_\C$-invariant, we see that as $\ger g'$-modules, we have 
\[
  L(\lambda)=\C[V]^{G_\C}\cong\C[X]^{G_\C}\cong\C[V_\R]^{G_\C}\Gamma.
\]
Thus, the latter is a copy of $L(\lambda)$ in $\Sw0{V_\R}^G\subseteq\Sw0{V_\R}$. Since $\Abs0{Dv}$ is invariant under the compact group $G$, it is determined by its restriction to $\Sw0{V_\R}^G$, which is determined by its values on $\C[V_\R]^G\Gamma\cong L(\lambda)$, due to the density of $\C[V_\R]\Gamma\subseteq\Sw0{V_\R}$. In particular, the latter restriction is non-zero. Computing on compactly supported superfunctions $\Gamma_c(\sh O_{V_\R})\subseteq\Sw0{V_\R}$, the following proposition readily follows.

\begin{Prop}
  The functional $\Abs0{Dv}$ is $\ger k$-relatively invariant for the character $-\lambda$. Hence, its restriction to $L(\lambda)\cong\C[V_\R]^G\Gamma$ spans the space $\GHom[_\ger k]0{L(\lambda),\C_{-\lambda}}$.
\end{Prop}


\section{The Riesz superdistribution}

In this section, we introduce a certain quadratic morphism $Q$, which pushes the Berezin integration functional $\Abs0{Dv}$ forward to a superdistribution on (an integration cycle in) $\GEnd0{\C^{p|q}}$. Moreover, we define the Riesz superdistribution $R_n$ and show in two different ways that it equals the pushforward of $\Abs0{Dv}$. 

\subsection{The super-Grassmannian and the integration cycle $\Omega$}

  Consider the complex super-Grassmannian $Y\defi\Gr_{p|q,2p|2q}(\C)$. Following Ref.~\cite{manin-gauge} with minor modifications, it is given as follows: Given a subset $I\subseteq 2p|2q$ of $p|q$ homogeneous indices, let $U_I$ be the superdomain with $S$-valued points
  \begin{equation}\label{eq:gr-localcoord}
    R_I=\kbordermatrix{~ & p|q \cr
      K & Z_I  \cr
      I & 1 },
  \end{equation}
  where $I$ indexes the $p|q$ rows containing the identity matrix and $K$ indexes the other $p|q$ rows, which make up the matrix $Z_I$. In other words, $U_I=\C^{p|q\times p|q}$, with generic point given by taking $S$ to be $\C^{p|q\times p|q}$ in the above definition, and $Z_I$ to be the matrix standard coordinate superfunctions. 

  Given another set $J$ of $p|q$ indices, let $B_{JI}$ be the $p|q\times p|q$ submatrix of $R_I$ formed by the rows indexed by $J$. Then $U_{IJ}$ is defined to be the maximal open subdomain of $U_I$ on which $B_{IJ}$ is invertible. The equation 
  \[
    Z_J=Z_IB_{JI}^{-1}
  \]
  expresses the entries of $Z_J$ as rational functions of the entries of $Z_I$, and defines an automorphism of the complex supermanifold $U_{IJ}$. Then $Y=\Gr_{p|q,2p|2q}(\C)$ is defined to be the complex supermanifold obtained by gluing these data. We identify $W\defi\C^{p|q\times p|q}$ with the standard affine open patch $U_{I_0}$, $I_0=\{p+1,\dots,2p|q+1,\dots,2q\}$.
  
  If $S$ is a complex superdomain, then the set of $S$-valued points in $\Gr_{p|q,2p|2q}(\C)$ is the set of equivalence classes $[x]$ of even $2p|2q\times p|q$ matrices $x$ with entries in $\Gamma(\sh O_S)$ such that the left $\Gamma(O_S)$-module $\Gamma(\sh O_S)^{2p|2q\times 2p|2q}\cdot x$ is projective of rank $p|q$; the equivalence relation identifies $x$ with $y$ if and only if the corresponding maps of right multiplication by these matrices have the same kernel.

  \subsubsection*{Supergroup actions}

  Consider the complex Lie supergroup $G'_\C\defi\GL(2p|2q,\C)$ whose Lie superalgebra is $\ger g'$. In the sequel we will write $g\in_S G'_\C$ (for any \emph{cs} manifold $S$) in the form
  \begin{equation}\label{eq:gprimeelts}
    \kbordermatrix{~ & p|q & p|q \cr
    p|q & A & B  \cr
    p|q & C & D}.  
  \end{equation}
  Then $G'_\C$ acts transitively on $\Gr_{p|q,2p|2q}(\C)$ by left multiplication. For $g\in_SG'_\C$ given in the form above and $Z\in_S W$, we have 
  \begin{equation}\label{partial-action}
    g\cdot Z = (AZ+B)(CZ+D)^{-1}\in_SW,
  \end{equation}
  whenever $CZ+D\in_S\GL(p|q,\C)$. In particular, the action of the complex supergroup $K_\C\defi\GL(p|q,\C)\times\GL(p|q,\C)$, realised as a closed subsupergroup of $G'_\C$ by requiring $B=C=0$ in the above notation, leaves the affine patch $W\subseteq Y$ invariant. The same is true for the closed subsupergroup $P^+$ of $G'_\C$ whose $S$-valued points are 
  \[
    \kbordermatrix{ ~ & p|q & p|q \cr
            p|q & 1 & B \cr
            p|q & 0 & 1 }.
  \]
  Consider the closed Lie subsupergroup $P^-$ of $G'_\C$ whose $S$-valued points are 
  \[
    \kbordermatrix{ ~ & p|q & p|q \cr
            p|q & 1 & 0 \cr
            p|q & C & 1 }.
  \]
  Then for $o_0\in W\subseteq Y$, the isotropy subsupergroup is
  \[
    G'_{\C,o_0}=K_\C P^-,
  \]
  which intersects trivially with $P^+$. In particular, $P^+$ acts simply transitively on $W$.

  We define a \emph{cs} form $H$ of $K_\C$ by specifying the real form $H_0\subseteq K_{\C,0}$ to be $\GL(p,\C)\times\mathrm U(q)$, embedded into $K_{\C,0}$ as the set of all matrices of the form 
  \[
  \kbordermatrix{
     & p & q & p & q \\
    p & A & &  & \\
    q & & D & & \\
    p & & & (A^*)^{-1} & \\
    q& & & & D'}
  \]
  with $A\in \GL(p,\C)$, $D,D'\in \mathrm{U}(q)$. Since $H$ is a closed subsupergroup of $K_{\C,cs}$, the orbit $\Omega\defi H.o_1$, where $o_1$ is the identity matrix in $W$, is a closed \emph{cs} submanifold of $W_{cs}$. The isotropy supergroup $H_o$ is the intersection (fibre product) of the diagonal subsupergroup $\GL(p|q,\C)_{cs}$ with $H$. In particular, we have 
  \[
    H_{o_1,0}=\mathrm U(p)\times\mathrm U(q), 
  \]
  embedded diagonally into $H$, and
  \[
    \dim_{cs}\Omega=p^2+q^2|2pq,\quad\Omega_0\cong \Herm^+(p)\times\mathrm U(q).
  \]

\subsection{The $Q$ morphism}

We let a quadratic map $Q:V\to W$ be defined as
\begin{equation}
  Q(L,L')\defi LL',\quad L\in\C^{p|q\times n},L'\in\C^{n\times p|q}.
\end{equation}
It is clearly $G_\C$-invariant, \cf Equation \eqref{eqn:gprimespo}. It gives rise to a corresponding morphism of complex supermanifolds. Moreover, defining a \emph{cs} form $W_\R$ of $W$ by setting
\[
  W_{\R,\ev}\defi\Herm(p)\times\Herm(q),
\]
one readily sees that $Q$ descends to a morphism $V_\R\to W_\R$ of \emph{cs} manifolds. The following is fairly straightforward.

\begin{Prop}
  The pullback along the morphism $Q$ induces a continuous linear map $Q^\sharp:\Sw0{W_\R}\to\Sw0{V_\R}$. In fact, for $Q^\sharp(f)$ to lie in $\Sw0{V_\R}$, it is sufficient for $f\in\Gamma(\sh O_{W_\R})$ to have rapid decay at infinity along $\Herm^+(p)$, \ie 
  \[
    \sup\nolimits_{z\in\Herm^+(p)}\Abs1{(1+\Norm0z)^N(Df)(z,w)}<\infty
  \]
  for all $w\in\Herm(q)$, $N\in\N$, and $D\in S(W)$. Here, $Df$ denotes the natural action of $S(W)$ by constant coefficient differential operators.

  In particular, $Q_\#(\Abs0{Dv})$, defined by 
  \[
    \Dual0{Q_\sharp(\Abs0{Dv})}f\defi\Dual0{\Abs0{Dv}}{Q^\sharp(f)},\quad f\in\Sw0{W_\R},
  \]
  is a continuous linear functional on $\Sw0{W_\R}$ with support in $\overline{\Herm^+(p)}$. Thus, it extends to a continuous functional on the space of all superfunctions $f\in\Gamma(\sh O_{W_\R})$ with rapid decay along $\Herm^+(p)$.
\end{Prop}

Moreover, note that we have 
\begin{equation}\label{eq:qfinitevects}
  Q^\sharp(\C[W_\R]e^{-\str})=\C[V_\R]^{G_\C}\Gamma,
\end{equation}
since the $G_\C$-invariants of $\C[V_\R]$ are generated in degree two, \cf Ref.~\cite{lsz}. 

There is an action of a suitable twofold cover $\tilde H$ of the \emph{cs} form $H$ of $K_\C$ on the space $\Sw0{V_\R}$. Explicitly, it can be written for any $\tilde h\in_S\tilde H$ lying above $h=\diag(A,D)\in_SH$, $f\in\Sw0{V_\R}$, and $(L,L')\in_SV_\R$, as
\[
  (\tilde h\cdot f)(L,L')=\Ber0A^{n/2}\Ber0D^{-n/2}f(D^{-1}L,L'A).
\]

Under the pullback $Q^\sharp$, this corresponds to the \emph{twisted action} $\cdot_\lambda$ of $\tilde H$ on $\Sw0{W_\R}$, 
\begin{equation}\label{eq:twact}
  (\tilde h\cdot_\lambda f)(w)=\Ber0A^{n/2}\Ber0D^{-n/2}f(D^{-1}wA).
\end{equation}
The \emph{untwisted action} of $H$ is defined to be 
\begin{equation}\label{eq:untwact}
  (h\cdot f)(w)\defi f(D^{-1}wA).
\end{equation}
The subspaces $\C[V_\R]\Gamma$ and $\C[W_\R]e^{-\str/2}$ are invariant for the induced $\ger k$-action.

\subsection{Statement of the theorem}

In what follows, recall the facts on Berezin integration summarised in the Appendix. 

The homogeneous \emph{cs} manifold $\Omega=H.o_1=H/H_{o_1}$ is a locally closed \emph{cs} submanifold of $W_{cs}$. It admits a non-zero $H$-invariant Berezinian density $\Abs0{Dy}$ \cite{ah-berezin}. 

Explicitly, it is given as follows. Observe that $\Omega$ has purely even codimension in $W_{cs}$. Thus, we have a canonical splitting $\Omega\cong\Omega_0\times W_\odd$, defining a retraction $r$ of $\Omega$, which we call \emph{standard}. Moreover, the standard coordinates of $W_{cs}$, \viz
\[
  Z=\kbordermatrix{%
      & p & q \\
    p & z & \zeta\\
    q & \omega & w
  }
\]
restrict to superfunctions $z_{ij},w_{k\ell},\zeta_{i\ell},\omega_{kj}$ on $\Omega$. For any local system $(x_a)$ of coordinates on $\Omega_0$, $(r^\sharp(x_a),\zeta,\omega)$ is a local system of coordinates on $\Omega$. 

In particular, $D(\zeta,\omega)$ is a well-defined relative Berezinian (density) on $\Omega$ over $\Omega_0$, with respect to the standard retraction. Denote by $\Abs0{dz}$ the Lebesgue density on $\Herm(p)$, and by $\Abs0{dw}$ the normalised invariant density on $\mathrm U(q)$. We set 
\[
  \Abs0{Dy}\defi D\mu(Z)\defi\frac{\Abs0{dz}\,\Abs0{dw}}{\Abs0{\det z}^p}\,D(\zeta,\omega)\,\det(z-\zeta w^{-1}\omega)^q\det(w-\omega z^{-1}\zeta)^p.
\]
Then $\Abs0{Dy}$ is the up to constants unique invariant Berezinian density on $\Omega$ \cite{as-sbos}.

\subsubsection*{Riesz superdistribution}

When $n\geqslant p$, define the functional $T_n$, called the \Define{Riesz superdistribution}, by 
\[
  \Dual0{T_n}f\defi\int_\Omega \Abs0{Dy}\,\Ber0y^n f(y)
\]
for any entire superfunction $f\in\Gamma(\sh O_W)$, which satisfies Paley--Wiener type estimates along the tube $T_0\defi\Herm^+(p)+i\Herm(p)$, \ie 
\begin{equation}\label{eq:pw-tube}
  \sup\nolimits_{z\in T_0}\Abs1{e^{-R\Norm0{\Im z}}(1+\Norm0z)^N(Df)(z,w)}<\infty
\end{equation}
for any $D\in S(W)$, $N\in\N$, $w\in\C^{q\times q}$, and some $R>0$. The integral is taken with respect to the standard retraction, and its convergence is proved in Ref.~\cite{as-sbos}.

Our terminology is explained by the fact that for $q=0$, $T_n$ coincides with the unweighted \emph{Riesz distribution} for the parameter $n$, see Ref.~\cite{FK94}. Using the super Laplace transform and some Functional Analysis, one proves the following \cite{as-sbos}.

\begin{Prop}[riesztemp]
  Let $n\geqslant p$. Then the functional $T_n$ extends continuously to the space of all superfunctions of rapid decay along $\Herm^+(p)$.
\end{Prop}

\subsubsection*{Conical superfunctions and Gindikin $\Gamma$ function}

To state the superbosonisation identity, we introduce the following set of rational superfunctions on $W$. For any $Z=(Z_{ij})\in_S W=\ger{gl}(p|q,\C)$ and $1\sle k\sle p+q$, we consider the $k$th principal minor $[Z]_k$ of $Z$, \viz
\[
  [Z]_k\defi(Z_{ij})_{1\sle i,j\sle k}. 
\]
Whenever $[Z]_k$ is invertible, we set $\Delta_k(Z)\defi\Ber0{[Z]_k}$, and whenever all principal minors of $Z$ are invertible and $\mathbf m=(m_1,\dots,m_{p+q})\in\Z^{p+q}$, we define
\begin{equation}
  \Delta_\mathbf m\defi\Delta_1^{m_1-m_2}\cdots\Delta_{p+q-1}^{m_{p+q-1}-m_{p+q}}\Delta_{p+q}^{m_{p+q}}.
\end{equation}
These functions are called \emph{conical superfunctions}. They are characterised as the unique rational superfunctions that are eigenfunctions of a suitable Borel \cite{as-sbos}.

Fix a superfunction $f\in\Gamma(\sh O_\Omega)$ and $x\in_SW_{cs}$. Whenever the integral converges, we define the \emph{Laplace transform} of $f$ at $x$ by 
\[
  \LT(f)(x)\defi\int_\Omega |Dy|\,e^{-\str(xy)}f(y),
\]
where we write $\Abs0{Dy}$ for the invariant Berezinian $\mu$ on $\Omega$. All integrals will be taken with respect to the standard retraction on $\Omega$. In particular, provided the integral exists, we define
\begin{equation}
  \Gamma_\Omega(\mathbf m)\defi\LT(\Delta_\mathbf m)(1)=\int_\Omega |Dy| e^{-\str(y)}\Delta_\mathbf m(y),
\end{equation}
and call this the \emph{Gindikin $\Gamma$ function} of $\Omega$. The following is proved in Ref.~\cite{as-sbos}.

\begin{Prop}[ltcon]
Let $O\subseteq W_{cs}$ be the open \emph{cs} submanifold on which all principal minors of $Z$ are invertible. For $x\in_SO$, the integral
\[
  \LT(\Delta_\mathbf m)(x^{-1})=\int_\Omega |Dy|\,e^{-\str(x^{-1}y)}\Delta_\mathbf m(y)
\]
converges absolutely if and only if $m_j>j-1$ for $j=1,\dots,p$. In this case, $\Gamma_\Omega(\mathbf m)$ exists, and we have
\[
  \LT(\Delta_\mathbf m)(x^{-1})=\Gamma_\Omega(\mathbf m)\Delta_\mathbf m(x).
\]
\end{Prop}

In fact, the function $\Gamma_\Omega(\mathbf m)$ can be determined explicitly, as follows, \cf Ref.~\cite{as-sbos}.

\begin{Th}[gamma]
  Let $m_j>j-1$ for all $j=1,\dots,p$. We have
  \[
    \Gamma_\Omega({\mathbf m}) =(2\pi)^{\frac{p(p-1)}2}\prod_{j=1}^p \Gamma(m_j-j+1)
    \prod_{k=1}^q \frac{\Gamma(q-k+1)}{\Gamma(m_{p+k}+q-k+1)}\frac{\Gamma(m_{p+k}+k)}{\Gamma(m_{p+k}-p+k)}.
  \]
  In particular, $\Gamma_\Omega(\mathbf m)$ extends uniquely as a meromorphic function of $\mathbf m\in\C^{p+q}$, which has neither zeros nor poles provided that 
  \[
    m_j>j-1,\quad j=1,\dots,p,\quad m_{p+k}>p-k,\quad k=1,\dots,q.
  \]
\end{Th}

\subsubsection*{Superbosonisation identity}

We are finally in a position to state the superbosonisation identity. To that end, denote for $n\geqslant p$: 
\[
  \Gamma_\Omega(n)\defi\Gamma_\Omega(n,\dots,n)>0,
\]
and let $R_n\defi\Gamma_\Omega(n)^{-1}T_n$ be the \emph{normalised Riesz superdistribution}. Then we have the following theorem \cite{as-sbos,lsz}.

\begin{Th}[sbos]
  Let $n\geqslant p$. Then we have 
  \[
    Q_\sharp(\Abs0{Dv})=\sqrt\pi^{np}R_n.
  \]
  Explicitly, for any holomorphic superfunction $f$ on the open subspace of $W$ whose underlying open set is $T_0+\C^{q\times q}$, , satisfying the estimate in Equation \eqref{eq:pw-tube} for some $R>0$ and any $D\in S(W)$, $w\in\C^{q\times q}$, and $N\in\N$, we have 
  \[
    \int_{V_\R}\Abs0{Dv}\,f(Q(v))=\frac{\sqrt{\pi}^{np}}{\Gamma_\Omega(n)}\int_\Omega \Abs0{Dy}\,\Ber0y^n f(y).
  \]
  In particular, this applies to any $f$ in the space $\C[W_\R]e^{-\str/2}$ from Equation \eqref{eq:qfinitevects}.
\end{Th}

\section{Proofs of the superbosonisation identity}

We end this survey by explaining two proofs of the superbosonisation identity. The first proof, which we only sketch briefly, makes heavy use of Functional Analysis to reduce everything to a trivial computation. The second proof equates the Riesz superdistribution $R_n$ to a relatively invariant functional on a suitable representation of $\ger g'$ and uses some basic representation theory to prove the identity. 

\subsection{Analytic proof}

One can prove the superbosonisation identity (\ie \thmref{Th}{sbos}) by computing the super version of the Euclidean Laplace transform of both sides of the identity and comparing the results. Given a sufficiently well-developed Functional Analysis machinery, one shows that the Laplace transform is injective. Here, we give a brief sketch of the procedure; for a more detailed discussion, in particular, of the relevant locally convex topologies, see \cite[Appendix C]{as-sbos}.

The space of continuous linear functionals on $\Sw0{W_\R}$ is denoted by $\TDi0{W_\R}$. Its elements are called \emph{tempered superdistributions}. Clearly, $\TDi0{V_\R}$ embeds continuously as a subspace into $\Gamma_c(\sh O_{W_\R})'$. The elements in the image are characterised as those functionals on $\Gamma_c(\sh O_{W_\R})$ that are continuous for the topology induced by $\Sw0{V_\R}$. Let $\mu\in\TDi0{W_\R}$. One can show the following \cite{as-sbos}.

\begin{Prop}
  There exists a (unique) largest open subspace $\gamma^\circ_{\mathcal S'}(\mu)\subseteq W_\R$ such that for every \emph{cs} manifold $S$ and any $w\in_S\gamma^\circ_\mathcal S(\mu)$, we have 
  \[
    e^{-\str(w\cdot)}\mu\in\Gamma(\sh O_S)\widehat\otimes\TDi0{W_\R},
  \]
  where $\widehat\otimes$ denotes the completed tensor product (w.r.t.~the injective or, equivalently, the projective tensor product topology).
\end{Prop}

Let $z=x+iy\in_SW_{cs}$ where $y\in_S W_\R$ and $x\in_S\gamma^\circ_\mathcal S(\mu)$ (this does \emph{not} determine $x,y$ uniquely). Then we define the \emph{Laplace transform} of $\mu$ by 
\[
  \LT(\mu)(z)\defi\mathcal F(e^{-\str(x\cdot)}\mu)(y),
\]
where $\sh F$ denotes the Fourier transform. This definition makes sense, since by a straightforward extension of Schwartz's classical theory of the Laplace transform, we have $e^{-\str(x\cdot)}\mu\in\Gamma(\sh O_S)\widehat\otimes\Sw0{W_\R}$, and hence the Fourier transform (w.r.t. $W_\R$) is contained in the same space. The following is a special case of results from Ref.~\cite{as-sbos}.

\begin{Prop}
  The Laplace transform $\LT(\mu)(z)$ is the value at $z$ of a holomorphic superfunction $\LT(\mu)$ on the tube $W_\R+i\gamma_\mathcal S^\circ(\mu)$. The tempered superdistribution $\mu$ is uniquely determined by $\LT(\mu)$. 
\end{Prop}

We can now finally give an account of the functional analytic proof of the superbosonisation identity. 

\begin{proof}[Proof of \thmref{Th}{sbos}, analytic version]
  Let $T\subseteq W$ be the open subspace whose underlying open set is $T_0+\C^{q\times q}$, where we recall that $T_0\defi\Herm^+(p)+i\Herm(p)$. For any $z\in_ST$ such that all principal minors of $z$ are invertible, we have 
  \[
    \LT(T_n)(z)=\LT(\Delta_{n,\dots,n})(z)=\Gamma_\Omega(n)\Delta_{n,\dots,n}(z^{-1})=\Gamma_\Omega(n)\Ber0z^{-n},
  \]
  in view of \thmref{Prop}{ltcon}. On the other hand, computing Gaussian Berezin integrals over affine superspace gives 
  \[
    \LT(Q_\sharp(\Abs0{Dv}))(z)=\sqrt\pi^{np}\Ber0z^{-n},
  \]
  thereby proving the theorem.
\end{proof}

Note that in the above sketch of the proof, we have omitted one non-trivial technical detail (stated above, in \thmref{Prop}{riesztemp}), namely, the argument that shows that $T_n$ indeed extends as a continuous functional on $\Sw0{W_\R}$. In Ref.~\cite{as-sbos}, this is achieved by discussing the domain in which the Laplace transform of a functional on the Paley--Wiener space is holomorphic. This reduces the proof of the temperedness of $T_n$ to a careful convergence discussion of the integral defining its Laplace transform.

\subsection{Representation theoretic proof}

  We end our survey by an account of a representation theoretic proof of the superbosonisation identity. Let $G'$ be the \emph{cs} form $G'_\C$ with underlying Lie group $G'_0\defi\mathrm U(p,p)\times\mathrm U(2q)$, embedded into $G'_\C=\GL(2p|2q,\C)$ as block matrices
  \[
    \kbordermatrix{
      & p & q & p & q \\
      p & a & 0& b & 0\\
      q & 0& a' & 0& b'\\
      p & c & 0& d & 0\\
      q& 0& c' &0 & d'},
    \quad
    \begin{pmatrix}
      a&b\\ c&d
    \end{pmatrix}\in\mathrm U(p,p),\
    \begin{pmatrix}
      a'&b'\\
      c'&d'
    \end{pmatrix}\in\mathrm U(2q).
  \]
  Here, we follow the conventions of Equation \eqref{eq:gprimeelts}. 

\subsubsection*{Line bundle}
  
The $\ger k$-character $2\lambda$ integrates to a character $\chi_{2\lambda}$ of $K_\C$, namely 
\[
  \chi_{2\lambda}\begin{pmatrix}
    A & 0\\
    0 & D
    \end{pmatrix}=\Ber0D^n\Ber0A^{-n}
\]
Extending $\chi_{2\lambda}$ trivially to $K_\C P^-$, we may define a holomorphic line bundle on the complex homogeneous superspace $G'_\C/K_\C P^-=Y$ by
\[
  L_{2\lambda}\defi G'_\C \times^{K_\C P^-_\C} \chi_{2\lambda}.
\]
We will also consider its restriction to $D:=G'.o_0$, where $o_0\in W_0$ is the zero matrix. Observe that the underlying space of $D$ is
\[
  D_0=\mathrm U(p,p)/(\mathrm U(p)\times\mathrm U(p))\times\mathrm U(2q)/(\mathrm U(q)\times\mathrm U(q)),
\]
the direct product of the set of $p\times p$ complex matrices of operator norm less than one and the Grassmannian of complex $q$-planes in $2q$-space.

Then $G'_\C$ naturally acts on sections of $L_{2\lambda}$, and $G'$ acts on sections of $L_{2\lambda}|_D$. Denote the latter action by $\pi_{2\lambda}$.

On general grounds, the cocycle defining $L_{2\lambda}$ is $\chi_{2\lambda}(s_J^{-1}s_I)$ where $s_I:U_I\to G'_\C$ is a local section of the $K_\C P^-_\C$-principal bundle $G'_\C\to Y$. The transition matrix for the super-Grassmannian $Y$ on $U_{IJ}$ is given by $\smash{Z_J=Z_IB_{JI}^{-1}}$. We have 
\[
s_J^{-1}s_I
=\begin{pmatrix}
  1&0\\0&B_{JI}^{-1}
\end{pmatrix}
\begin{pmatrix}
1&Z\\0&1
\end{pmatrix}\in_{U_{IJ}}K_\C P^-_\C, 
\]
so that the defining cocycle of $L_{2\lambda}$ is 
\[
\chi_{2\lambda}(s_J^{-1}s_I)=\Ber0{B_{JI}}^n.
\]

\subsubsection*{Highest weight section}

We construct a global section $|0\rangle$ of the line bundle $L_{2\lambda}$ as follows: In the trivialisation on $U_I$ given by $s_I$, it is defined by
\[
|0\rangle_I(Z_I)=\Ber0{s_I(Z_I)}^{-n}.
\]
It is not difficult to see that this indeed defines a global section of $L_{2\lambda}$, which on the standard affine patch $W=U_{I_0}$ is just the constant function $1$.

Let $Z\in_SD\cap W$ and $g\in_SG'$ be such that $g\cdot Z\in_S D\cap W$, \ie the action remains in the affine patch. Then the action of $g$ on a section $f$ of $L_{2\lambda}|_D$ can be expressed at $Z$ by 
\[
\pi_{2\lambda}(g)f(Z)=\chi_{2\lambda}(k(g,Z))f(g^{-1}\cdot Z),
\]
where 
\[
k(g,Z)\defi\kbordermatrix{~ & p|q & p|q \cr
p|q & (A-BD^{-1}C)(1+ZD^{-1}C)^{-1} & 0  \cr
p|q & 0 & CZ+D}\in_SK_\C.
\]
This follows immediately from the relation 
\[
g\;
\begin{pmatrix}
  1&Z\\0&1
\end{pmatrix}
\in
\begin{pmatrix}
  1&g\cdot Z\\0&1
\end{pmatrix}
k(g,Z)P^-(S).
\]
Since holomorphic sections of $L_{2\lambda}$ are determined by their restriction to $W=U_{I_0}$, one computes that $|0\rangle$ is fixed by the action of $P^-$ and transforms under $K_\C$ by the character $\chi_{2\lambda}$. Therefore, there is a $\ger g'$-equivariant map
\[
  M(2\lambda)\defi\Uenv0{\ger g'}\otimes_{\ger k\oplus\ger p^+}\C_\lambda\to\Gamma(D,L_{2\lambda}|_D)
\]
from the parabolic Verma module of highest weight $2\lambda$, which maps the highest weight vector to $|0\rangle$. 

\subsubsection*{Construction of an intertwiner}

We define the \Define{weighted Laplace transform} $\LT_n$ by
\[
  \LT_n(f)(z)\defi\Gamma_\Omega(n)^{-1}\LT(f\Delta_{n,\dots,n})(z/2)=\Gamma_\Omega(n)^{-1}\int_\Omega\Abs0{Dy}e^{-\str(zy)/2}f(y)\Ber0y^n,
\]
whenever this makes sense. For $h\in_SH$, we have
\begin{equation}\label{eq:lapequi}
  \LT_n(h\cdot f)(z)=\chi_{2\lambda}(h)\LT_n(f)(h^{-1}\cdot z),
\end{equation}
where $h\cdot f$ denotes the untwisted $H$-action introduced in Equation \eqref{eq:untwact}.

The \emph{Cayley transform} is $\gamma(Z)\defi(1+Z)(1-Z)^{-1}$, and the \Define{weighted Cayley transform} $\gamma_n$ is defined by
\[
  \gamma_n(F)(Z)\defi \Ber0{1-Z}^{-n}F(\gamma(Z)),
\]
when this is well-defined. For the diagonal subsupergroup $K'_\C\defi\diag\GL(p|q,\C)$ of $K_\C$, $\gamma$ is $K_\C'$-equivariant. Thus, $\gamma_n\circ\LT_n$ intertwines the $\ger k'$-action by $\pi_n$ on $\Gamma(D,L_{2\lambda}|_D)$ and the untwisted $\ger k'$-action on $\C[W_\R]e^{-\str/2}$. 

By \thmref{Prop}{ltcon}, it follows easily that 
\[
  (\gamma_n\circ\LT_n)(e^{-\str/2})=1=|0\rangle|_{D\cap W}.
\]
More generally, we have 
\begin{equation}\label{eq:wtlapcon}
  (\gamma_n\circ\LT_n)(\Delta_\mathbf me^{-\str/2})(z)=(n)_\mathbf m\,\Delta_{\mathbf m+n}(1-z)
\end{equation}
for $\mathbf m\in\Z^{p+q}$ with $m_j>j-1$ for $j=1,\dots,p$. Here, $\mathbf m+n\defi(m_1+n,\dots)$ and 
\[
  (n)_\mathbf m\defi\Gamma_\Omega(n)^{-1}\Gamma_\Omega(\mathbf m+n).
\]

The superfunction $\Delta_\mathbf m$ is polynomial (\ie contained in $\C[W_\R]$) if and only if 
\begin{equation}\label{eq:mpol}
  m_1\geqslant m_2\geqslant\cdots\geqslant m_p\geqslant 0,\quad m_{p+1}\sle m_{p+2}\sle\cdots m_{p+q}\sle0.
\end{equation}
For these $\mathbf m$, we have $\Gamma_\Omega(\mathbf m+n)\neq0$ if and only if in addition $m_{p+1}\geqslant p-n$, as one sees by applying \thmref{Th}{gamma}. 

It is not hard to see that those $\Delta_\mathbf m$ that are polynomial are exactly the lowest weight vectors for the $K_\C$-action on $\C[W_\R]$ \cite{as-sbos}. Moreover, $\C[W_\R]$ is semi-simple (and multiplicity free), according to \cite[Proposition 3.1]{scheunert-zhang}. Thus, we have a $(\ger k,H)$-module decomposition
\[
  \C[W_\R]=\bigoplus\nolimits_\mathbf m\GEnd0{L^{p|q}(\mu_\mathbf m)},
\]
where $\mathbf m$ runs over all multi-indices satisfying the assumptions of Equation \eqref{eq:mpol}, 
\[
\GEnd0{L^{p|q}(\mu_\mathbf m)}\cong\C[W_\R]_\mathbf m\defi\Uenv0{\ger k}\Delta_\mathbf m, 
\]
and with respect to the standard Borel, $L^{p|q}(\mu_\mathbf m)$ is the simple finite-dimensional $\ger{gl}(p|q,\C)$-module of highest weight
\[
  \mu_\mathbf m\defi-\sum_{j=1}^pm_j\delta_j+\sum_{j=p+1}^{p+q}m_j\eps_{j-p}.
\]

By Equations \eqref{eq:lapequi} and \eqref{eq:wtlapcon}, and $\LT_n(fe^{-\str/2})(z)=\LT_n(f)(1-z)$, the summand $\C[W_\R]_\mathbf m$ is annihilated by $\LT_n$ if $m_{p+1}<n-p$ and mapped injectively otherwise. Since $e^{-\str/2}$ is invariant under $K_\C'$, the composite $\gamma_n\circ\LT_n$ is a split epimorphism of $\ger k'$-modules from $\C[W_\R]e^{-\str/2}$ onto its image. 

We can now give a proof of the superbosonisation identity (\ie \thmref{Th}{sbos}) by Representation Theory.

\begin{proof}[Proof of \thmref{Th}{sbos}, representation theoretic version]
We consider the untwisted $\ger k$-action on $\C[W_\R]e^{-\str/2}$. By the invariance of $\Abs0{Dy}$, the restriction of $R_n$ defines an element of $\GHom[_\ger k]0{\C[W_\R]e^{-\str/2},\C_{-2\lambda}}$. Since
\[
  Q^\sharp:\C[W_\R]e^{-\str/2}\otimes\C_\lambda\to\C[V_\R]\Gamma
\]
is a surjective $\ger k$-equivariant map, we obtain an injection
\[
  Q_\sharp:\GHom[_\ger k]0{\C[V_\R]\Gamma,\C_{-\lambda}}\to\GHom[_\ger k]0{\C[W_\R]e^{-\str/2}\otimes\C_\lambda,\C_{-\lambda}}
\]
by left exactness of the hom functor $\GHom[_\ger k]0{\cdot,\cdot}$. The latter of these hom spaces is $\GHom[_\ger k]0{\C[W_\R]e^{-\str/2},\C_{-2\lambda}}$. Since $\gamma_n\circ\LT_n(e^{-\str/2})=|0\rangle$, after Cayley transformation, we obtain a parabolic of $\ger g'$ whose nilradical annihilates $\LT_n(e^{-\str/2})$. 

In particular, the dimension of $\GHom[_\ger k]0{\C[W_\R]e^{-\str/2},\C_{-2\lambda}}$ is at most one, and it is spanned by both $Q_\sharp(\Abs0{Dv})$ and $R_n$. Computing constants, the claim follows. 
\end{proof}
\section*{Appendix}
%
\subsection*{Supergeometry} We summarise the basic definitions from supergeometry we use throughout the text. 

\begin{Def}[super]
  A \emph{$\C$-superspace} is a pair $X=(X_0,\sh O_X)$ where $X_0$ is a topological space and $\sh O_X$ is a sheaf of supercommutative superalgebras over $\C$ with local stalks. A \emph{morphism} $f:X\to Y$ of $\C$-superspaces is a pair $(f_0,f^\sharp)$ comprising a continuous map $f_0:X_0\to Y_0$ and a sheaf map $f^\sharp:f_0^{-1}\sh O_Y\to O_X$, which is local in the sense that $f^\sharp(\ger m_{Y,f_0(x)})\subseteq\ger m_{X,x}$ for any $x$, where $\ger m_{X,x}$ is the maximal ideal of $\sh O_{X,x}$.

  Global sections $f\in\Gamma(\sh O_X)$ of $\sh O_X$ are called \Define{superfunctions}. Due to the locality condition, the \Define{value} $f(x)\defi f+\ger m_{X,x}\in\sh O_{X,x}/\ger m_{X,x}=\C$ is defined for any $x$. \Define{Open subspaces} of a $\C$-superspace $X$ are given by $(U,\sh O_X|_U)$, for any open subset $U\subseteq X_0$.
\end{Def}

We consider two types of model spaces.

\begin{Def}[model]
  For a complex super-vector space $V$, we define $\sh O_V\defi\sh H_{V_\ev}\otimes\bigwedge(V_\odd)^*$ where $\sh H$ denotes the sheaf of holomorphic functions. The space $(V_\ev,\sh O_V)$ is called the \Define{complex affine superspace} associated with $V$, and denoted by $V$.

  If instead, $V$ is a \emph{cs} vector space (see \thmref{Def}{csvs}), then we define the sheaf $\smash{\sh O_V\defi\sh C^\infty_{V_\ev}\otimes\bigwedge(V_\odd)^*}$, where $\sh C^\infty$ denotes the sheaf of complex-valued smooth functions. The space $(V_\ev,\sh O_V)$ is called the \Define{\emph{cs} affine superspace} associated with $V$, and denoted by $V$. (The \emph{cs} terminology is due to J.~Bernstein.)
\end{Def}

In turn, this gives two flavours of supermanifolds.

\begin{Def}[sman]
  Let $X$ be a $\C$-superspace, whose underlying topological space $X_0$ is Hausdorff, and which admits a cover by open subspaces isomorphic to open subspaces of some complex resp.~\emph{cs} affine superspace $V$, where $V$ may vary. Then $X$ is called a \Define{complex supermanifold} resp.~a \Define{\emph{cs} manifold}. 

  Complex supermanifolds and \emph{cs} manifolds form full subcategories of the category of $\C$-superspaces that admit finite products. The assignment sending the complex affine superspace $V$ to the \emph{cs} affine superspace obtained by forgetting the complex structure on $V_\ev$ extends to a product-preserving \emph{\emph{cs}-ification} functor from complex supermanifolds to \emph{cs} manifolds; the \emph{cs}-ification of $X$ is denoted by $X_{cs}$.
\end{Def}

This point of view is also espoused by Witten in recent work \cite{witten}.

\subsection*{Supergroups and supergroup pairs} We give some basic definitions on supergroups. Details can be found in \cite{manin-gauge,deligne-morgan,fioresi-bk}.

  \begin{Def}
    A \emph{complex Lie supergroup} (resp.~ a \emph{\emph{cs} Lie supergroup}) is a group object in the category of complex supermanifolds (resp.~\emph{cs} manifolds). A \emph{morphism} of (complex or \emph{cs}) Lie supergroups is a morphism of group objects in the category of complex supermanifolds (resp.~\emph{cs} manifolds). The \emph{cs}-ification functor maps complex Lie supergroups to \emph{cs} Lie supergroups and morphisms of complex Lie supergroups to morphisms of \emph{cs} Lie supergroups.
  \end{Def}

  \begin{Def}
    A complex (resp.~\emph{cs}) \emph{supergroup pair} $(\ger g,G_0)$ is given by a complex (resp.~real) Lie group $G_0$ and a complex Lie superalgebra $\ger g$, together with a morphism $\Ad:G_0\to\Aut0{\ger g}$ of complex (resp.~real) Lie groups such that $\ger g_\ev$ is the Lie algebra of $G_0$ (resp.~its complexification), $\Ad$ extends the adjoint action of $G_0$ on $\ger g_\ev$, and $[\cdot,\cdot]$ extends $d\Ad$. A \emph{morphism of supergroup pairs} $(d\phi,\phi_0)$ consists of a morphism $\phi_0$ of complex (resp.~real) Lie groups and a morphism $d\phi$ of Lie superalgebras that is $\phi_0$-equivariant for the $\Ad$-actions, such that $d\phi$ extends $d(\phi_0)$.
  \end{Def}

  The following is well-known, \cf Refs.~\cite{kostant,koszul,fioresi-bk}.

  \begin{Prop}
    There is an equivalence of the categories of complex (resp.~\emph{cs}) Lie supergroups and of complex (resp.~\emph{cs}) supergroup pairs. It maps any Lie supergroup to the pair consisting of its Lie superalgebra and its underlying Lie group.
  \end{Prop}

  \begin{Def}
    A closed embedding of (complex resp.~\emph{cs}) Lie supergroups is called a closed (complex resp.~\emph{cs}) \emph{subsupergroup}. A \emph{closed supergroup subpair} $(\ger h,H_0)\subseteq(\ger g,G_0)$ consists of a Lie subsuperalgebra $\ger h\subseteq\ger g$ and a closed subgroup $H_0\subseteq G_0$, such that $(\ger h,H_0)$ is a supergroup pair. Given a complex Lie supergroup $G$, a \emph{\emph{cs} form} of $G$ is a closed subsupergroup $H$ of $G_{cs}$ such that in the supergroup pairs $(\ger h,H_0)$ and $(\ger g,G_0)$ of $H$ resp.~$G$, one has $\ger h=\ger g$. In this case, $H_0$ is a real form of $G_0$.
  \end{Def}

  If $G$ is a complex Lie supergroup with associated supergroup pair $(\ger g,G_0)$, then $(\ger g,H_0)$, for a closed subgroup $H_0\subseteq G_0$, is the supergroup pair of a \emph{cs} form of $G$ if and only if $H_0$ is a real form of $G_0$, or equivalently, if $(\ger g,H_0)$ is a \emph{cs} supergroup pair. To define a \emph{cs} form $H$ of $G$, it thus suffices to specify a real form $H_0\subseteq G_0$.

\subsection*{Points}

If $\cat C$ is any category, and $X$ is an object of $\cat C$, then an \Define{$S$-valued point} (where $S$ is another object of $\cat C$) is defined to be a morphism $x:S\to X$. One may view this as a `parametrised' point. Suggestively, one writes $x\in_SX$ in this case, and denotes the set of all $x\in_SX$ by $X(S)$.

For any morphism $f:X\to Y$, one may define a set-map $f_S:X(S)\to Y(S)$ by 
\[
  f_S(x)\defi f(x)\defi f\circ x\in_SY,\quad x\in_SX.
\]
Clearly, the values $f(x)$ completely determine $f$, as can be seen by evaluating at the \Define{generic point} $x=\id_X\in_XX$.

In fact, more is true. The following statement is known as Yoneda's Lemma: Given a collection of set-maps $f_S:X(S)\to Y(S)$, there exists a morphism $f:X\to Y$ such that $f_S(x)=f(x)$ for all $x\in_SX$ if and only if 
\[
  f_T(x(t))=f_S(x)(t),\quad t:T\to S.
\]
The points $x(t)$ are called \Define{specialisations of $x$}, so the condition states that the collection $(f_S)$ is invariant under specialisation.

The above facts are usually stated in the following more abstract form: For any object $X$, $X(-):\cat C^{op}\to\cat{Sets}$ is a set-valued functor, and the set of natural transformations $X(-)\to Y(-)$ is naturally bijective to the set of morphisms $X\to Y$. Thus, the \Define{Yoneda embedding} $X\mapsto X(-)$ from $\cat C$ to $[\cat C^{op},\cat{Sets}]$, is fully faithful.

The Yoneda embedding preserves products, so if $\cat C$ admits finite products, it induces a fully faithful embedding of the category of group objects in $\cat C$ into the category $[\cat C^{op},\cat{Grp}]$ of group-valued functors. In other words, we have the following: Let $X$ be an object in $\cat C$. Then $X$ is a group object if and only if for any $S$, $X(S)$ admits a group law that is invariant under specialisation.

\subsection*{Berezin integrals}

Let $X$ be a \emph{cs} manifold and $\sh Ber_X$ to the Berezinian sheaf. The sheaf of \Define{Berezinian densities} $\Abs0{\sh Ber}_X$ is the twist by the orientation sheaf. Given local coordinates $(x^a)=(x,\xi)$ on $U$, one may consider the distinguished basis 
\[
  \Abs0{D(x^a)}=\Abs0{D(x,\xi)}=dx_1\dots dx_p\frac{\partial^\Pi}{\partial\xi^1}\dots\frac{\partial^\Pi}{\partial\xi^q}
\]
of the module of Berezinian densities $\Abs0{\sh Ber}_X$ \cite{manin-gauge}.

A \Define{retraction} of $X$ is a morphism $r:X\to X_0$ that is left inverse to the canonical embedding $j:X_0\to X$. A system of coordinates $(x,\xi)$ of $X$ is called \Define{adapted} to $r$ if $x=r^\sharp(x_0)$. Given such an adapted system, we may write $\omega=\Abs0{D(x,\xi)}\,f$ and 
\[
  f=\sum_{I\subseteq\{1,\dots,q\}}r^\sharp(f_I)\,\xi^I
\]
for unique coefficients $f_I\in\Gamma(\sh O_{X_0})$, where $\dim Y=*|q$. Then one defines
\[
  \fint_{X/X_0}\omega\defi\Abs0{dx_0}\,f_{\{1,\dots,q\}},
\]
which is an ordinary density on $X_0$. This quantity only depends on $r$, and not on the choice of an adapted system of coordinates. 

If the resulting density is absolutely integrable on $X_0$, then we say that $\omega$ is \Define{absolutely integrable} with respect to $r$, and define
\[
  \int_X\omega\defi\int_{X_0}{}\fint_{X/X_0}\omega.
\]
Unless $\supp\omega$ is compact, this quantity and its existence depend heavily on $r$.

\end{document}